\documentclass[aps,a4paper,superscriptaddress,nofootinbib,twocolumn]{revtex4-1}
\usepackage{amsmath,amssymb,gensymb,graphicx,multirow,dcolumn,bm,latexsym,soul,nicefrac,acronym}
\usepackage[mathscr]{euscript}
\usepackage{appendix}
\usepackage[colorlinks,linkcolor=blue,citecolor=blue,urlcolor=blue ]{hyperref}
\usepackage[normalem]{ulem}
\usepackage{float}
\usepackage{soul}
\usepackage{verbatim}
\usepackage{color}
\usepackage{acronym}
\usepackage{subfigure}
\usepackage{appendix}
\usepackage{longtable}
\usepackage{footnote}
\usepackage{cancel}
\usepackage{ulem}

\newcommand{\lzy}[1]{\textcolor{blue}{ZY: #1}}

\newcommand{\TRC}{MOE Key Laboratory of TianQin Mission, TianQin Research Center for Gravitational Physics $\&$  School of Physics and Astronomy, Frontiers Science Center for TianQin, CNSA Research Center for Gravitational Waves, Sun Yat-sen University (Zhuhai Campus), Zhuhai 519082, China}

\newcommand{\HNU}{School of Physics, Henan Normal University, Xinxiang 453007, China}
\begin{document}

\title{Mapping Anisotropies in the Stochastic Gravitational-Wave Background with space detector networks}
\author{Zhi-Yuan Li}
\affiliation{\TRC}
\author{Zheng-Cheng Liang}
\email{Corresponding author: liangzhengcheng@htu.edu.cn}
\affiliation{\HNU}
\author{Cong-mao Zhang}
\affiliation{\TRC}
\author{Jian-dong Zhang}
\affiliation{\TRC}
\author{Yi-Ming Hu}
\email{Corresponding author: huyiming@sysu.edu.cn}
\affiliation{\TRC}

\date{\today}

\begin{abstract}

Future space-based gravitational-wave detectors such as TianQin, LISA, and Taiji are expected to conduct joint observations. Such a multi-detector network will provide complementary viewing angles for the anisotropic stochastic gravitational-wave background (SGWB), thereby significantly enhancing the capability to reconstruct and localize its spatial distribution.
In this paper, we have established the first dedicated data analysis pipeline for the anisotropic stochastic gravitational-wave background using a joint network of TianQin, LISA, and Taiji. 
Our analysis incorporates both Gaussian, stationary, and unpolarized point sources from diverse sky locations as well as a random sky map. 
We have performed full-sky map reconstruction in pixel space using maximum likelihood estimation to extract the angular distribution of the SGWB. 
The results demonstrate that, when considering the detector noise, the TianQin+LISA+Taiji detector network can reconstruct the angular power spectrum of the stochastic background up to a maximum multipole moment of $l = 14 $, which can provide valuable information for studies on the spatial distribution of galactic compact binaries and physical imprints from the early Universe.
\end{abstract}

\maketitle
\acrodef{SGWB}{stochastic gravitational wave background}
\acrodef{GW}{gravitational-wave}
\acrodef{CBC}{compact binary coalescence}
\acrodef{MBHB}{supermassive black hole binary}
\acrodef{BBH}{binary black hole}
\acrodef{BNS}{binary neutron star}
\acrodef{EMRI}{extreme-mass-ratio inspiral}
\acrodef{DWD}{double white dwarf}
\acrodef{GDWD}{Galactic double white dwarf}
\acrodef{EDWD}{extragalactic double white dwarf}
\acrodef{BH}{black hole}
\acrodef{NS}{neutron star}
\acrodef{BNS}{binary neutron star}
\acrodef{LIGO}{Laser Interferometer Gravitational-Wave Observatory}
\acrodef{LISA}{Laser Interferometer Space Antenna}
\acrodef{TQ}{TianQin}
\acrodef{KAGRA}{Kamioka Gravitational Wave Detector}
\acrodef{ET}{Einstein telescope}
\acrodef{DECIGO}{DECi-hertz interferometry GravitationalWave Observatory}
\acrodef{CE}{Cosmic Explorer}
\acrodef{NANOGrav}{The North American Nanohertz Observatory for Gravitational Waves}
\acrodef{LHS}{left-hand side}
\acrodef{RHS}{right-hand side}
\acrodef{ORF}{overlap reduction function}
\acrodef{ASD}{amplitude spectral density}
\acrodef{PSD}{power spectral density}
\acrodef{SNR}{signal-to-noise ratio}
\acrodef{FIM}{Fisher information matrix}
\acrodef{TDI}{time delay interferometry}
\acrodef{PIS}{peak-integrated sensitivity}
\acrodef{PT}{phase transitions}
\acrodef{PLIS}{power-law integrated sensitivity}
\acrodef{GR}{general relativity}
\acrodef{PBH}{primordial black hole}
\acrodef{SSB}{solar system baryo}
\acrodef{PTA}{pulsar timing arrays}
\acrodef{SVD}{singular-value decomposition}
\acrodef{JSD}{Jensen-Shannon divergence}
\acrodef{CSD}{cross-spectral density}
\acrodef{TJ}{Taiji}
\acrodef{CMB}{cosmic microwave background}

\section{Introduction}
The \ac{SGWB} arises from the superposition of numerous unresolved \ac{GW} sources, originating from both astrophysical and cosmological processes. It offers a unique window into the physics of the early universe, and provides insights into astrophysical populations that are inaccessible through individually resolved sources ~\cite{Regimbau:2011rp,Caprini:2018mtu,Cusin:2018rsq,Capurri:2021zli,Bellomo:2021mer}. 
The astrophysical origins of the \ac{SGWB} include binary systems such as \ac{DWD}~\cite{Huang:2020rjf,Korol:2017qcx}, \ac{BBH}~\cite{Wang:2019ryf,Chen:2023qga,Liu_2020,Wang:2023tle}, \ac{BNS}~\cite{Surace:2015ppq,Talukder:2014eba}, and \ac{EMRI}~\cite{Fan:2020zhy,Zi:2021pdp,Ye:2023uvh}, among others. On the other hand, cosmological sources include early-universe phenomena such as inflation~\cite{Guth:1982ec}, first-order \acp{PT}~\cite{Hogan:1983ixn}, and cosmic defects~\cite{Kibble:1976sj,LISACosmologyWorkingGroup:2022jok}.

Stochastic gravitational waves are indistinguishable from unidentified instrumental noise in a single detector. To overcome this challenge, methods such as cross-correlation~\cite{Hellings:1983fr,Allen:1997ad,LIGOScientific:2017zlf,Liang:2021bde,Liang:2022ufy,Liang:2023fdf,Wang:2022pav,Wang:2023ltz,Hu:2024toa} and null channels~\cite{Tinto:2001ii,Tinto:2001ui,Hogan:2001jn,Robinson:2008fb,Romano:2016dpx,Smith:2019wny,Muratore:2021uqj} have been proposed. 
The cross-correlation method enables the extraction of the SGWB signal by correlating data from multiple detectors under the assumption that the noise in different detectors is uncorrelated. 
By employing this method, both ground-based interferometers and \acp{PTA} have made remarkable progress. Ground-based detectors have constrained the dimensionless energy density of the SGWB in the hundreds of Hertz~\cite{KAGRA:2021kbb}, whereas \acp{PTA} have recently reported compelling evidence for the presence of a stochastic background in the nano-hertz (nHz)~\cite{NANOGrav:2023gor,Xu:2023wog,EPTA:2023fyk,Reardon:2023gzh}.
The null-channel method is specifically proposed for future space-based gravitational wave detectors such as \ac{TQ}~\cite{TianQin:2015yph}, \ac{LISA}~\cite{LISA:2017pwj}, and \ac{TJ}~\cite{Hu:2017mde,Ruan:2018tsw}, which operate in the milli-Hertz (mHz) band. It aims to address the challenge of detecting the \ac{SGWB} when only a single detector is available. 
These detectors typically consist of three satellites with six links. By constructing signal-insensitive null channels as noise-monitoring references, one can effectively subtract correlated noise from the auto-correlation channels and thereby extract the \ac{SGWB}.
Most of the earlier studies assume a single detector, where the only viable choice would be the null channel.
However, the null channel method has to adopt precise knowledge of the instrumental noise \emph{a priori}.
\cite{Cornish:2001hg,Taruya:2005yf,Contaldi:2020rht,Li:2024lvt,Banagiri:2021ovv,Renzini:2022alw,Rieck:2023pej,Bloom:2024ffn}. 
On the other hand, since the space-borne \ac{GW} missions are expected to be launched around the same time, it is possible to perform joint observation by forming a network of detectors.
In this scenario, it is possible to perform cross correlation with multiple detectors and gain a model-independent estimate of the \ac{SGWB}~\cite{Liang:2021bde,Liang:2022ufy,Wang:2022pav,Liang:2023fdf,Wang:2023ltz,Hu:2024toa,Zhao:2024yau,Mu:2025gtg,Cheng:2025kku}.

Some sources of the \ac{SGWB} may give rise to potential anisotropies, characterized by spatial variations in its intensity~\cite{Wang:2019ryf,Huang:2020rjf,Liu:2020eko,Fan:2020zhy,Siemens:2006yp,Figueroa:2012kw,Mandic:2016lcn,Saikawa:2018rcs,Caprini:2018mtu,Auclair:2019wcv,Wang:2020jrd,Cui:2023dlo}. 
The anisotropy of the SGWB can open a new window into the study of large-scale structure~\cite{Cusin:2018rsq,Capurri:2021zli,Bellomo:2021mer} and the invaluable primordial information from the early Universe~\cite{Geller:2018mwu,Jenkins:2018nty,Li:2021iva,Wang:2021djr,Profumo:2023ybp,Schulze:2023ich}. 
While the null-channel method shows limited sensitivity to anisotropic \ac{SGWB}, it has been suggested that the cross-correlation method can greatly enhance their sensitivity to anisotropic features of the SGWB~\cite{Liang:2023fdf,Zhao:2024yau,Mentasti:2023uyi}.
However, a complete data analysis pipeline for joint detection has not yet been fully established. In this work, we aim to develop a data processing pipeline for anisotropic \ac{SGWB} using a network composed of TianQin, LISA, and Taiji.

The process of extracting directional information of the \ac{SGWB} from the data is referred to as ``map-making". It was originally developed in the context of the \ac{CMB}~\cite{Knox:1995dq,Bond:1997wr,Zaldarriaga:1997ch,Bond:1998zw} and later adapted for use in studies of the \ac{SGWB}~\cite{Allen:1996gp,Cornish:2001hg,Kudoh:2005as,Ungarelli:2001xu,Cornish:2002bh,Seto:2004np,Mitra:2007mc}. 
A widely adopted technique is the maximum likelihood method~\cite{Talukder:2010yd,Renzini:2018vkx,Suresh:2020khz,Agarwal:2021gvz,Renzini:2021iim,Xiao:2022uvq}, which retrieves the sky localization of the stochastic signal—referred to as the clean map—by deconvolving the detector response from the observed data, commonly known as the dirty map. 
It has also been widely applied to individual space-based gravitational wave detectors~\cite{Contaldi:2020rht,Li:2024lvt}. 
Under idealized noise conditions, LISA is expected to reconstruct the spherical harmonic components of the SGWB anisotropy up to multipole moments $l<7$~\cite{Contaldi:2020rht}, while TianQin is restricted to $l<4$, and additionally suffers from symmetry-induced degeneracies along its orbital plane~\cite{Li:2024lvt}. 
Beyond the maximum likelihood framework, the Bayesian spherical harmonic method offers a promising alternative by avoiding the deconvolution instability. However, this comes at the expense of substantially increased computational complexity~\cite{Banagiri:2021ovv,Renzini:2022alw,Rieck:2023pej,Bloom:2024ffn,Criswell:2024hfn}.

In this paper, we implement the map-making procedure for the joint TianQin–LISA–Taiji network using the maximum likelihood estimation method. We simulate stationary, Gaussian, unpolarized point-source signals and a random skymap.
Under the stationary Gaussian noise conditions, we demonstrate the ability of the detector network to reconstruct sky maps for signals with different spectral shapes.

The manuscript is organized as follows.
Section~\ref{sec:basic} reviews the formalism of \ac{SGWB} detection, while Sec.~\ref{sec:method} details the maximum likelihood method and the associated deconvolution procedure for map-making. 
In~Sec.~\ref{sec:simulation}, we describe the procedure for generating simulated data and highlight important considerations in the process. 
Based on the previous sections, we present the results of map-making with different spectral shapes in Sec.~\ref{sec:result}. Finally, we give a summary in Sec.~\ref{sec:summary}.
\section{Formalism}\label{sec:basic}
\subsection{The signal of stochastic background}\label{sec:signal}
The \ac{SGWB} arises from the superposition of a large number of plane waves emitted from different directions. In transverse-traceless gauge, the metric perturbations $h(t,\vec{x})$ associated with the \ac{SGWB} can be written as:
\begin{equation}
\begin{split}
    h_{ij}(t,\vec{x}) = &\sum_{P = +, \times} 
    \int ^{\infty}_{-\infty}{\rm d}f 
    \int_{S^2} {\rm d}\hat{\Omega}_{\hat{\mathbf{n}}}\,
    e_{ij}^P(\hat{\mathbf{n}})\widetilde{h}_P(f,\hat{\mathbf{n}}) \\
    &\times {\rm e}^{{\rm i} 2\pi f [ t+\hat{\mathbf{n}}\cdot \vec{x}(t)/c ]},
\end{split}
\end{equation}
where the $\vec{x}(t)$ denotes the position of the measurement, $e_{ij}^P(\hat{\mathbf{n}})$ is the \ac{GW} polarization tensor corresponding to the polarization $P=+, \times$, and $c$ denotes the light speed. 
The Fourier amplitude $\widetilde{h}_P(f,\hat{\mathbf{n}})$ captures the frequency-domain characteristics of \ac{GW} for each specific propagation direction $\hat{\mathbf{n}}$. 
Assuming the \ac{SGWB} is {\it anisotropic}, Gaussian, and unpolarized 
from astrophysical origins, the \ac{SGWB} at this stage is formed by the superposition of numerous independent astronomical signals. 
The Gaussian assumption is validated by the central limit theorem. 
Therefore, the Fourier amplitude $\widetilde{h}_P(f,\hat{\mathbf{n}})$ follows a zero-mean Gaussian distribution, with a variance of
\begin{equation}
	\label{eq:hh}
    \langle \widetilde{h}_P(f,\hat{\mathbf{n}})\widetilde{h}^{\ast}_{P'}(f',\hat{\mathbf{n}}') \rangle = \frac{1}{4}\delta(f-f')\delta_{PP'}\delta^2(\hat{\mathbf{n}}-\hat{\mathbf{n}}')\mathscr{P}_{\rm h}(f,\hat{\mathbf{n}}),
\end{equation}
$\mathscr{P}_{\rm h}(f,\hat{\mathbf{n}})$ is the one-sided \ac{PSD}, the unpolarized property guarantees that $\mathscr{P}_{\rm h}(f,\hat{\mathbf{n}}) = \mathscr{P}_{\rm +}(f,\hat{\mathbf{n}}) = \mathscr{P}_{\rm \times}(f,\hat{\mathbf{n}}) = 1/2 \mathscr{P}_{\rm GW}(f,\hat{\mathbf{n}})$. Note that the one-sided \ac{PSD} is defined to be twice the two-sided \ac{PSD}; hence, the factor of 1/4 arises from both the unpolarized assumption and the use of the one-sided convention.

Without considering the Doppler effect, one can assume that the direction and the intensity of \ac{SGWB} are independent,
\begin{equation}
	\label{eq:Ph}
    \mathscr{P}_{\rm h}(f,\hat{\mathbf{n}}) = \bar{H}(f)\mathcal{P}_{\rm h}(\hat{\mathbf{n}}),
\end{equation}
where $\bar{H}(f)$ is the \ac{SGWB} spectral shapes.
The angular distribution $\mathcal{P}_{\rm h}(\hat{\mathbf{n}})$ can be expanded in terms
of a set of bases, usually a spherical harmonic basis or a pixel basis.
On a pixel basis,
\begin{equation}
    \label{eq:Plm}
    \mathcal{P}_{\rm h}(\hat{\mathbf{n}}) = \mathcal{P}_{\hat{\mathbf{n}}'}\delta(\hat{\mathbf{n}},\hat{\mathbf{n}}'),
\end{equation}
on a spherical harmonic basis, it has the form:
\begin{equation}
    \label{eq:Plm}
    \mathcal{P}_{\rm h}(\hat{\mathbf{n}}) = \sum_{l = 0}^{\infty}\sum_{m = -l}^{l} p_{lm}Y_{lm}(\hat{\mathbf{n}}),
\end{equation}
Furthermore, in the spherical harmonic basis, one can define the angular power spectrum,
\begin{eqnarray}
        C_{l} &=& \frac{1}{2l+1}\sum_{m = -l}^{l}\left|\int_{S^2}p_{lm}Y_{lm}(\hat{\mathbf{n}})\frac{{\rm {d}}\hat{\mathbf{n}}}{4\pi}\right|^2\\ \nonumber
        &= &\frac{1}{2l+1}\sum_{m = -l}^{l}\left|p_{lm}\right|^2
\end{eqnarray}
To obtain the all-sky PSD of \ac{SGWB}, one needs to integrated the $\mathscr{P}_{\rm h}(f,\hat{\mathbf{n}})$ over all directions:
\begin{equation}
    \label{eq:S_h}
    S_{h}(f) = \bar{H}(f)\int_{S^2} {\rm d}\Omega_{\hat{\mathbf{n}}} \mathcal{P}_{\rm h}(\hat{\mathbf{n}}).
\end{equation}

Moreover, to characterize the energy distribution across different frequencies in the \ac{SGWB}, the dimensionless energy spectral density, $\Omega_{\rm gw}$, is commonly employed,
\begin{equation}
    \Omega_{\rm gw}(f)= \frac{1}{\rho_c}\frac{\rm{d}\rho_{\rm gw}}{{\rm d}(\ln f)},
\end{equation}
where $\rho_c = 3H^2_0c^2/{8\pi} G$ is the critical energy density with the gravitational constant G, and the Hubble constant $H_0$. The term ${\rm d}\rho_{\rm gw}$ represents the \ac{GW} energy density contained within the frequency interval $[f,f+df]$. 
The energy density spectrum $\Omega_{\rm gw}(f)$ can be related to the gravitational-wave strain \ac{PSD} $S_{h}(f)$:
\begin{equation}
    \Omega_{\rm gw}(f)= \frac{2\pi^2 f^3}{3H^2_0}S_{h}(f).
\end{equation}
Theoretical models predict a large number of energy density spectral shapes, among which the power-law spectrum is one of the most widely considered scenarios for its simplicity.
\begin{equation}
\label{eq:spectrum index}
    \Omega_{\rm gw}(f) 
    = \Omega_{\rm r}\left(\frac{f}{f_{\rm r}}\right)^{\alpha},
\end{equation}
where $\alpha$ is the power-law index,  $\Omega_{r}$ denotes the reference energy density at reference frequency $f_{\rm r}$, and the choice of reference frequency is arbitrary.

\subsection{Detector orbits}

TianQin is a space-borne gravitational wave detector consisting of three identical satellites in Earth orbit. The satellites follow a circular orbit with a radius of approximately $10^5$ km and an orbital period of 3.64 days.
In operation, the three satellites form an equilateral triangle with arm length approximately $1.7 \times 10^5$ km, and the normal vector of their orbital plane is oriented toward J0806($\theta_s = -4.7^{\circ},\phi_s = 120.5^{\circ}$).
TianQin operates in a ``three months on, three months off" mode, allowing for a cumulative observation time of six months each year~\cite{Hu:2018yqb}.

On the other hand, \ac{LISA} and Taiji are designed to orbit the Sun with a period of one year. 
Both of them adopt an equilateral triangular configuration. 
LISA features an arm length of $2.5\times10^6$ km, trails the Earth by~$20^\circ$, and its orbital plane is inclined at~$+60^\circ$ concerning the ecliptic. 
Taiji is proposed to be a LISA-like mission with a $3 \times 10^{6}$ km arm length. It has three candidate orbits Taijim, Taijip, and Taijic. In this paper, we adopt the Taijip orbit, which leads the Earth by $20^\circ$, with a~$+60^{\circ}$ inclined orientation~\cite{Li:2023szq}.
For convenience, we will use ``LS" to represent LISA, ``TQ" to represent TianQin, and ``TJ" to represent Taiji in the following equations and figures.

Due to differences in operational modes, the joint working time of multiple detectors also varies. TQ+LS and TQ+TJ can conduct joint observations for six months each year (3+3 months), while TJ+LS are capable of continuous joint observation throughout the entire year.

\subsection{Detector response and noise}
For space-based detectors, due to the detector motion, their response continuously changes over time. 
Typically, we divide the data into short segments, ensuring that the detector's response remains relatively stable and that the stationarity assumption holds for each data segment $[t_{0}-T/2,t_{0}+T/2]$, which is commonly referred to as the short-term Fourier transform. 
In the time domain, the \ac{SGWB} signal $h_I(f)$ in the detector channel $I$ can be approximated as 
\begin{eqnarray}
\label{eq:ht_sgwb}
\nonumber
h_{I}(t) &\approx&
\int_{-\infty}^{\infty}{\rm d}f \, \sum_{P}
\int_{S^{2}}{\rm d}\hat{\Omega}_{\hat{\mathbf{n}}}
F_{I}^{P}|_{t_{0}}(f,\hat{\mathbf{n}})\widetilde{h}_{P}(f,\hat{\mathbf{n}})\\
&&\times 
e^{{\rm i}2\pi f[t+\hat{\mathbf{n}}\cdot\vec{x}_{I}(t_{0})/c]},
\end{eqnarray}
where $h_P$ is the gravitational wave signal, the $\vec{x}_I(t_{0})$ denotes the position of the measurement at time $t_{0}$, the frequency-domain response $F_{I}^{P}|_{t_{0}}$ is determined by the polarization tensor and the detector channel $I$~\cite{Cornish:2001qi}. 
Notice that, for computational convenience, we approximate the time-domain response to the middle point of the time period.
We will explain the validity of the approximation in Sec. \ref{sec:segment}.
Furthermore, the \ac{SGWB} signal in frequency domain during the period $[t_{0}-T/2,t_{0}+T/2]$ can be written as:
\begin{equation}
	\label{eq:hf}
	\widetilde{h}_{I}(f)
	=\sum_{P}\int_{S^{2}}\,{\rm{d}}\hat{\Omega}_{\hat{\mathbf{n}}}\,
	F_{I}^{P}|_{t_{0}}(f,\hat{\mathbf{n}})\widetilde{h}_{P}(f,\hat{\mathbf{n}}) 
	\,{\rm e}^{{\rm i}2\pi f\hat{\mathbf{n}}\cdot\vec{x}(t_{0})/c},
\end{equation}
The \ac{PSD} and \ac{CSD} of the frequency-domain signal $\langle\widetilde{h}_{I}(f)\widetilde{h}_{J}(f)\rangle$ are determined by
\begin{equation}
    \label{eq: h_{IJ}}
    \langle\widetilde{h}_{I}(f)\widetilde{h}_{J}(f)\rangle = \frac{1}{2}\delta(f-f') \int_{S^2} {\rm d} \Omega_{\hat{\mathbf{n}}} \gamma_{IJ}|_{t_{0}} (f,\hat{\mathbf{n}}) \mathscr{P}_{\rm h}(f,\hat{\mathbf{n}}),
\end{equation}
where $\gamma_{IJ}|_{t_{0}} (f,\hat{\mathbf{n}})$ is the geometric factor that takes into account the separation and relative orientation of the two detectors~\cite{Liang:2023fdf,Thrane:2009fp,Finn:2008vh},
\begin{equation}
    \label{eq: gamma}
	\gamma_{IJ}|_{t_{0}}(f,\hat{\mathbf{n}})=
	\frac{1}{2} 
	\sum_{P} F_I^P|_{t_{0}}(f,\hat{\mathbf{n}}) F_{J}^{P*}|_{t_{0}}(f,\hat{\mathbf{n}})\,
	{\rm e}^{{\rm i} 2 \pi f \hat{\mathbf{n}} \Delta\vec{x}/c},
\end{equation}

where the factor of $1/2$ arises from the average of polarization.  $\Delta\vec{x} = \vec{x}_{I}-\vec{x}_{J}$ denotes the separation vector
between channels $I$ and $J$. 

The detection of an \ac{SGWB} depends not only on the detector response but also on the instrumental noise. 
Due to the motion of space-borne detectors, phase noise becomes the dominant noise source in gravitational wave detection. 
By applying time shifts and linear combinations of the measurement channels, an equivalent equal-arm Michelson interferometer can be constructed. 
This significantly suppresses the phase noise by several orders of magnitude, which is known as \ac{TDI}~\cite{Tinto:2022zmf,Tinto:2002de,Wang:2022nea,Muratore:2021uqj,Otto:2015erp}. 
Among the various TDI combinations, a commonly used choice is X/Y/Z channels~\cite{Vallisneri:2007xa}, each channel is centered on one satellite, with two laser beams transmitted through the four adjacent links to form two closed loops. These signals are then subtracted to effectively cancel out the laser phase noise, more details can be found in~\cite{Li:2023szq}. Under the assumption of equal arms
and equal noise levels, by performing linear combinations of the X/Y/Z channels, one can obtain the mutually orthogonal A/E/T channels~\cite{Adams:2010vc},
\begin{equation}
	\begin{split}
	\label{eq:AET}
	{\rm A}&=\frac{1}{\sqrt{2}}({\rm Z}-{\rm X}),\\
	{\rm E}&=\frac{1}{\sqrt{6}}({\rm X}-2{\rm Y}+{\rm Z}),\\
	{\rm T}&=\frac{1}{\sqrt{3}}({\rm X}+{\rm Y}+{\rm Z}).
\end{split}
\end{equation}
It is worth noting that at low frequencies, the T channel is insensitive to \ac{GW} signals and is therefore often used as a noise monitoring detector.
Assuming that the noise is stationary, the noise \ac{PSD} can be written as: 
\begin{equation}
    \begin{aligned}
    N_{{\mathrm{A} / \mathrm{E}}}(f)= & \frac{2 \sin ^2 u}{L^2}\left[(\cos u+2) S_{\mathrm{p}}(f)\right. \\
    & \left.+2(\cos (2 u)+2 \cos u+3) \frac{S_{\mathrm{a}}(f)}{(2 \pi f)^4}\right] \\
    N_{{\mathrm{T}}}(f)= & \frac{8 \sin ^2 u \sin ^2 \frac{u}{2}}{L^2}\left[S_{\mathrm{p}}(f)+4 \sin ^2 \frac{u}{2} \frac{S_{\mathrm{a}}(f)}{(2 \pi f)^4}\right],
\end{aligned}
\end{equation}
where $u = 2\pi fL/c$, which is determined by the detector arm length $L$. $S_{\mathrm{p}}$ and $S_{\mathrm{a}}$ represent position measurement noise and acceleration noise. For TianQin, LISA, and Taiji, comprehensive details about each detector parameter can be found in~\cite{TianQin:2020hid,Babak:2021mhe,Luo:2019zal}.

\begin{figure}
    \centering
    \includegraphics[width=0.5\textwidth]{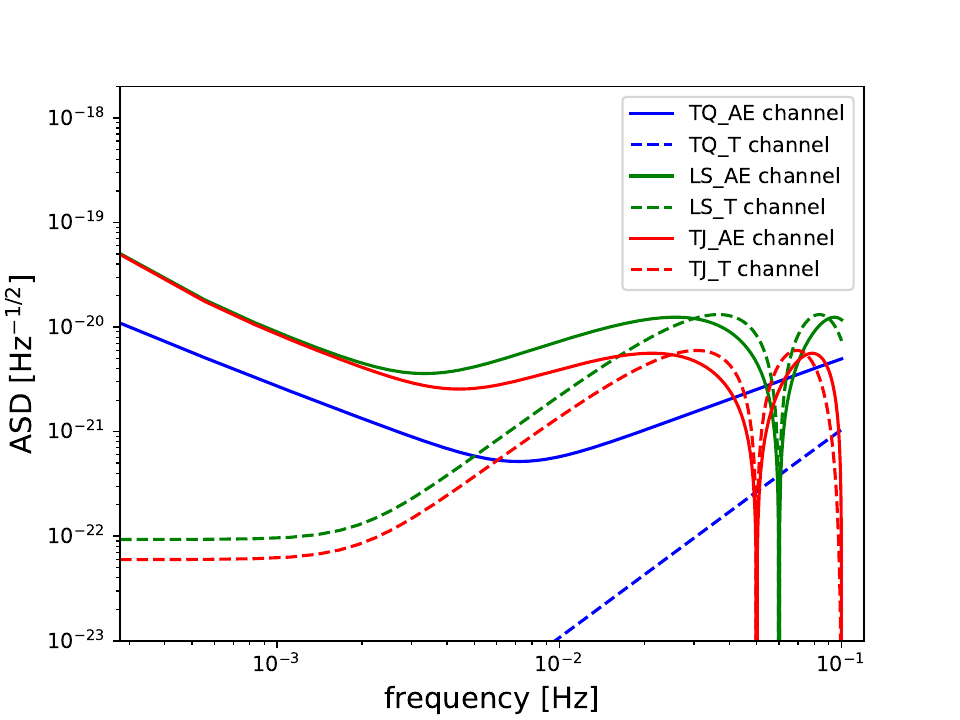}
    \caption{The noise \ac{PSD} for the A, E, and T channels of different detectors, where the A and E channels exhibit identical noise levels}
    \label{fig:noise PSD}
\end{figure}

Fig.~\ref{fig:noise PSD} displays the noise \ac{PSD} for the A, E, and T channels of different detectors, where the A and E channels exhibit identical noise levels.


\section{Map-making analysis}\label{sec:method}
In this section, we will construct a detection statistic from the data 
$\widetilde{d}_I$, and estimate the directional parameters of the signal using the maximum likelihood method.
Unlike a single space-borne detector, a detector network does not require prior information about the noise model~\cite{Contaldi:2020rht,Li:2024lvt}, which significantly enhances the robustness of the method. The overall approach is more similar to that of ground-based detectors~\cite{Thrane:2009fp,Renzini:2018vkx,Suresh:2020khz,Agarwal:2021gvz,Renzini:2021iim,Xiao:2022uvq}.

The two distinct basis representations for sky localization - pixelization and spherical harmonics introduced earlier are fundamentally equivalent in the map-making procedure.
In this work, we employ the HEALPix pixelization scheme~\cite{Gorski:2004by} as our foundational basis. 
HEALPix is an equal-area pixelization scheme for spherical surfaces, which discretizes the sky into uniformly sized pixels with identical surface areas, optimizing both computational efficiency and statistical rigor.
Unless otherwise specified, all results are presented in the pixel basis representation by default.
\subsection{Maximum likelihood method}
The data $d_I$ consist of the signal after response and the noise,
\begin{equation}
    \widetilde{d}_I(f) = \widetilde{h}_{I}(f) +  \widetilde{n}_{I}(f).
\end{equation}
Given the zero-mean Gaussian assumptions for both the noise $\widetilde{n}_{I}$ and the signal $\widetilde{h}_{I}$, the data $\widetilde{d}_I$ are likewise zero-mean Gaussian distributed. Therefore, we adopt the correlation measurement as the detection statistic for data processing.
\begin{equation}
    C_{IJ}|_{t_0}(f) = \frac{2}{\tau}\widetilde{d}_{I}|_{t_0}(f)\widetilde{d}^*_{J}|_{t_0}(f),
\end{equation}
where the factor 2 is consistent with the definition of one-sided PSD, $\tau$ denotes the duration of a segment after the short-time Fourier transform. Unlike the multi-channel configuration of a single space-borne detector, the indices $I$ and $J$ here correspond to distinct detectors within the network. Given the extremely large separation between the two detectors, we consider their noise to be statistically independent~\cite{PismaZhETF.23.326,Detweiler:1979wn,10.1093/mnras/227.4.933,Flanagan:1993ix,Christensen:1992wi}. 
The expectation of the correlation measurement $C_{IJ}$ is entirely determined by the signal, 
\begin{equation}
    \label{eq:C_IJ}
    \mu_{IJ}|_{t_0}(f) = \langle C_{IJ}|_{t_0}(f) \rangle = \bar{H}(f)\int {\rm d}^2\,\hat{\Omega}_{\hat{\mathbf{n}}}\gamma_{I J}|_{t_0}(f, \hat{\mathbf{n}})\mathcal{P}_{\rm h}(\hat{\mathbf{n}}).
\end{equation}
Assuming the signal is sufficiently weak compared to the noise, the covariance matrix is given by~\cite{Thrane:2009fp,Romano:2016dpx}:
\begin{equation}
	\begin{aligned}
		\label{eq:sigma_IJ}
		\sigma^{2}_{IJ}|_{t_0}(f)
		=&
		\langle C^{2}_{IJ}|_{t_0}(f) \rangle-\langle C_{IJ}|_{t_0}(f) \rangle^{2}\\
        \approx& N_{II}(f)N_{JJ}(f).
	\end{aligned}
\end{equation}
For strong signals, however, a correction to the covariance matrix is required. The explicit form can be found in~\cite{Liang:2022ufy,Li:2024lvt}.

We observe that the covariance matrix $\sigma_{IJ}^2$ is completely determined by the correlation $C_{IJ}$ and is independent of any assumptions about the noise model. Furthermore, the spectral shape $\bar{H}(f)$ can be readily estimated from the data~\cite{Adams:2010vc,Cheng:2022vct,Wang:2021njt,Liang:2024tgn}. Consequently, in our subsequent analysis, we treat both the covariance matrix $\sigma_{IJ}^2$ and the spectral shape $\bar{H}(f)$ as known quantities~\cite{Li:2025eog}, eliminating the need for any prior assumptions about the noise model.

In terms of Eq.~(\ref{eq:C_IJ}), the \ac{SNR} is expressed as follows~\cite{Liang:2023fdf},
\begin{equation}
	\label{eq:snr_tot}
	\rho_{IJ}=
	\sqrt{\sum_{f,t}
	\left(\frac{\mu_{IJ}|_{t}(f)}
	{\sigma_{IJ}|_{t}(f)}\right)^2}.
\end{equation}
For a detector network comprising multiple channels $\{IJ\}$ and baselines $\mathcal{N}$, the combined signal-to-noise ratio (SNR) is given by
\begin{equation}
    \rho_{\rm tot} = \sqrt{\sum_{\mathcal{N}}\sum_{IJ}\rho^2_{IJ}}
\end{equation}
where $\mathcal{N}$ is the number of detector network baselines, such as \ac{TQ}+\ac{LISA}, \ac{TQ}+\ac{TJ} and \ac{LISA}+ \ac{TJ}.

Using $\mathcal{P_{\hat{\mathbf{n}}}}$ denotes the  angular distribution in Eq.~(\ref{eq:Ph}). 
The likelihood can be written as~\cite{Thrane:2009fp}:
\begin{equation}
    \label{eq:llh_f}
    p(\{C_{ft}\}|\{\mathcal{P_{\hat{\mathbf{n}}}}\}) \propto {\rm exp}\left[-\frac{1}{2}\chi^2(\mathcal{P})\right],
\end{equation}
where 
\begin{equation}
\label{eq:chi-squre}
\begin{aligned}
    \chi^2(\mathcal{P}) = & \sum_{t}\sum_{f}(C_{IJ}^*|_t(f)-\bar{H}(f)\gamma_{IJ}^*|_t(f,\hat{\mathbf{n}})\mathcal{P}_{\hat{\mathbf{n}}}^*) \\
    & \frac{1}{\sigma_{IJ}^2|_t(f)} (C_{IJ}|_t(f)-\bar{H}(f)\gamma_{IJ}|_t(f,\hat{\mathbf{n}})\mathcal{P}_{\hat{\mathbf{n}}})
\end{aligned}
\end{equation}
Since maximizing the likelihood for $\mathcal{P}_{\hat{\mathbf{n}}}$ is equivalent to minimizing $\chi^2$, the maximum likelihood estimators for the $\mathcal{P}_{\hat{\mathbf{n}}}$ is given by:
\begin{equation}
	\label{eq:Pa}
	\hat{\mathcal{P}}_{\hat{\mathbf{n}}}
	=(F^{-1})^{IJ}_{\hat{\mathbf{n}}\hat{\mathbf{n}}'}X^{IJ}_{\hat{\mathbf{n}}'},
\end{equation}
The quantity $X^{IJ}_{\hat{\mathbf{n}}'}$, commonly known as the dirty map, characterizes the sky signal convolved with the detector response.  
To recover the underlying sky signal, one must invert the \ac{FIM} $F^{IJ}_{\hat{\mathbf{n}}\hat{\mathbf{n}}'}$
, yielding the components of the clean map $\hat{\mathcal{P}}_{\hat{\mathbf{n}}}$. This transformation from the dirty map to the clean map constitutes a deconvolution process, effectively disentangling the true astrophysical signal from instrumental effects. The explicit forms of $X^{IJ}_{\hat{\mathbf{n}}'}$ and $F^{IJ}_{\hat{\mathbf{n}}\hat{\mathbf{n}}'}$ are given by:
\begin{equation}
	\label{eq:X_IJ}
	X^{IJ}_{\hat{\mathbf{n}}'}=
	\sum_{f,t}\gamma^{*}_{IJ}|_t(f,\hat{\mathbf{n}}')
	\frac{\bar{H}(f)}{\sigma^{2}_{IJ}|_t(f)}C_{IJ}|_t(f),
\end{equation}

\begin{equation}
\label{eq:Fisher}
	F^{IJ}_{\hat{\mathbf{n}}\hat{\mathbf{n}}'}
	=\sum_{f,t}
	\gamma_{IJ}|_t(f,\hat{\mathbf{n}})\frac{\bar{H}^{2}(f)}{\sigma^{2}_{IJ}|_t(f)}
	\gamma^{*}_{IJ}|_t(f,\hat{\mathbf{n}}').
\end{equation}
In addition, one can show in the weak-signal approximation that:
\begin{equation}
    \langle X_{\hat{\mathbf{n}}} X^*_{\hat{\mathbf{n}}'}\rangle - \langle X_{\hat{\mathbf{n}}}\rangle \langle X^*_{\hat{\mathbf{n}}'}\rangle \approx F_{\hat{\mathbf{n}}\hat{\mathbf{n}}'},
\end{equation}
\begin{equation}
    \langle \hat{\mathcal{P}}_{\hat{\mathbf{n}}} \hat{\mathcal{P}}^*_{\hat{\mathbf{n}}'}\rangle - \langle \hat{\mathcal{P}}_{\hat{\mathbf{n}}}\rangle \langle \hat{\mathcal{P}}^*_{\hat{\mathbf{n}}'}\rangle \approx (F^{-1})_{\hat{\mathbf{n}}\hat{\mathbf{n}}'}.
\end{equation}
Therefore, $F_{\hat{\mathbf{n}}\hat{\mathbf{n}}'}$ is the covariance matrix of the dirty map $X_{\hat{\mathbf{n}}}$, and $(F^{-1})_{\hat{\mathbf{n}}\hat{\mathbf{n}}'}$ is the covariance matrix of the clean map $\hat{\mathcal{P}}_{\hat{\mathbf{n}}}$.

For multiple channel pairs $\{IJ\}$ and detector network $\mathcal{N}$, one can simply adds the dirty maps $X^{IJ}_{\hat{\mathbf{n}}'}$ and the $F^{IJ}_{\hat{\mathbf{n}}\hat{\mathbf{n}}'}$~\cite{Thrane:2009fp}, 
\begin{equation}
    X^{\rm tot}_{\hat{\mathbf{n}}'} = \sum_{\mathcal{N}}\sum_{IJ} X^{IJ}_{\hat{\mathbf{n}}'},
\end{equation}
\begin{equation}
    F^{\rm tot}_{\hat{\mathbf{n}}\hat{\mathbf{n}}'} = \sum_{\mathcal{N}}\sum_{IJ} F^{IJ}_{\hat{\mathbf{n}}\hat{\mathbf{n}}'}.
\end{equation}
This result is obtained by extending the likelihood formulation in Eq.~(\ref{eq:chi-squre}) to include sums over baselines as well as frequency and time. The maximum-likelihood estimator $\mathcal{P}_{\hat{\mathbf{n}}}$ has a form similar to that of the single-baseline case,
\begin{equation}
	\label{eq:Ptot}
	\hat{\mathcal{P}}_{\hat{\mathbf{n}}}
	=(F^{-1})^{\rm tot}_{\hat{\mathbf{n}}\hat{\mathbf{n}}'}X^{\rm tot}_{\hat{\mathbf{n}}'}.
\end{equation}

\subsection{Deconvolution and regularization}
Equation~(\ref{eq:Pa}) and~(\ref{eq:Ptot}) estimate the clean map by removing the effects of the point spread function, a process called deconvolution. The deconvolution requires computing the inverse of the \ac{FIM}, 
which is typically highly singular. Since single baselines and multiple baselines share \ac{FIM} of identical form, we uniformly denote the \ac{FIM} as $F_{\hat{\mathbf{n}}\hat{\mathbf{n}}'}$ throughout this section unless otherwise specified. 

The Fisher matrix becomes singular mainly due to two factors: the detector pair is diffraction-limited, and certain blind directions inherent in the detector geometry. 
The diffraction limit applies mostly to ground-based detectors, while for space-borne gravitational wave detectors, due to the very large separation between satellites, the diffraction limit can be neglected. 
Regarding the second factor, it is fundamentally determined by the detector's orbital configuration. 
For single space-based gravitational wave detectors, this issue severely constrains their sky map recovery capability~\cite{Li:2024lvt,Contaldi:2020rht}. However, joint observations with multiple detectors can effectively mitigate this limitation~\cite{Liang:2023fdf}. To address the singularity issue of the \ac{FIM} $F_{\hat{\mathbf{n}}\hat{\mathbf{n}}'}$ , we typically employ \ac{SVD}, since $F_{\hat{\mathbf{n}}\hat{\mathbf{n}}'}$ is Hermitian, its SVD has the form,
\begin{equation}
    F_{\hat{\mathbf{n}}\hat{\mathbf{n}}'} = U\Sigma U^*,
\end{equation}
where $U$ is a unitary matrix and $\Sigma$ is a diagonal matrix
containing positive real eigenvalues $s_i$ of \ac{FIM}. 
After performing SVD, we sort the singular values in descending order $s_1 \geq s_2\geq s_3...\geq s_n$. The smallest singular values, when inverted, introduce the most significant numerical errors. A standard regularization approach involves setting a threshold $s_{\rm min}$, where singular values below this cutoff are either replaced with infinity (whose inverse turns to zero) or clamped to the threshold value itself ($s_i = s_{\rm min}$) to maintain numerical stability~\cite{Thrane:2009fp,Li:2024lvt}.
Using this modified matrix $\Sigma'$ we can then define the regularized $F'_{\hat{\mathbf{n}}\hat{\mathbf{n}}'}$ as:
\begin{equation}
    F'_{\hat{\mathbf{n}}\hat{\mathbf{n}}'} = U\Sigma' U^*,
\end{equation}
with its inverse:
\begin{equation}
    F'^{-1}_{\hat{\mathbf{n}}\hat{\mathbf{n}}'} = U\Sigma'^{-1} U^*.
\end{equation}

\begin{figure}
    \centering
    \includegraphics[width=0.5\textwidth]{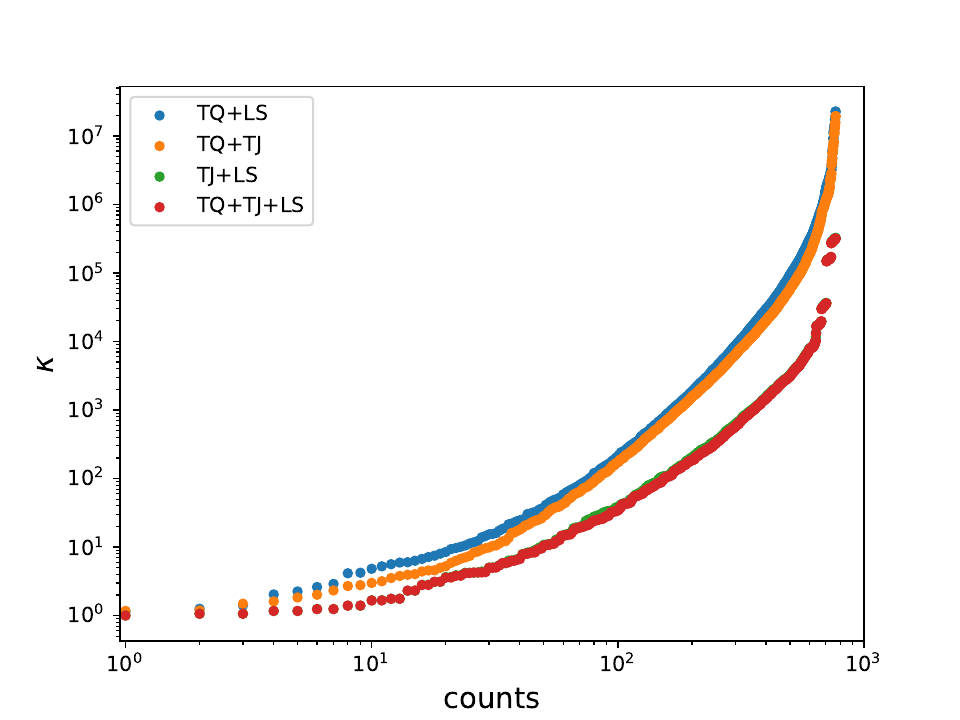}
    \caption{The condition numbers for various detector network configurations (TianQin, Taiji, and LISA in different combinations). The value of “counts” represents how many leading eigenvalues we retain. The spectral index in Eq.~(\ref{eq:spectrum index}) is set to $\alpha = 2/3$.}
    \label{fig:FIM condition number}
\end{figure}

A more intuitive approach to characterize matrix singularity is through the condition number $\kappa$~\cite{Agarwal:2021gvz,kincaid2009numerical}, which quantifies the sensitivity of the matrix to numerical errors. It is defined as the ratio of the largest to the smallest eigenvalue, 
\begin{equation}
    \kappa(F_{\hat{\mathbf{n}}\hat{\mathbf{n}}'}) = \lambda_{\rm max}/\lambda_{\rm min},
\end{equation}
where $\lambda_{\rm max}$ and $\lambda_{\rm min}$ represent the maximum and minimum eigenvalues of the $F_{\hat{\mathbf{n}}\hat{\mathbf{n}}'}$, respectively. A larger condition number indicates a more singular matrix, which introduces greater numerical errors during matrix inversion. 
The condition number of the \ac{FIM} is influenced by multiple factors, including: the spectral shape of gravitational wave signals,
the orbital configuration of detectors, and the chosen frequency band for analysis~\cite{Li:2024lvt,Contaldi:2020rht}. Fig.~\ref{fig:FIM condition number} shows the progressive condition number (that is, the condition number for incrementally larger submatrix) of different space detector baselines. 
Here we adopt the healpix parameter $N_{\rm side} = 8$, which partitions the sky map into 768 pixels, and corresponds to 768 eigenvalues in the \ac{FIM}.
The value of “counts” represents how many
leading eigenvalues we retain. Compared to a single detector, a detector network can reduce the matrix condition number by over 10 orders of magnitude~\cite{Li:2024lvt}, with the TJ+LS configuration demonstrating the most significant improvement. 

In addition, after using the regularized \ac{FIM} $F'_{\hat{\mathbf{n}}\hat{\mathbf{n}}'}$, the clean map in Eq.~(\ref{eq:Pa}) and (\ref{eq:Ptot}) can be replaced by
\begin{equation}
    \hat{\mathcal{P}}'_{\hat{\mathbf{n}}} = F'^{-1}_{\hat{\mathbf{n}}\hat{\mathbf{n}}'} X_{\hat{\mathbf{n}}},
\end{equation}
and the expectation and covariance of $\mathcal{P}'_{\hat{\mathbf{n}}}$ have also been changed, 
\begin{equation}
    \langle \hat{\mathcal{P}}'_{\hat{\mathbf{n}}}\rangle = F'^{-1}F\mathcal{P}_{\hat{\mathbf{n}}} \neq \mathcal{P}_{\hat{\mathbf{n}}}
\end{equation}
\begin{equation}
    \langle \hat{\mathcal{P}}'_{\hat{\mathbf{n}}} \hat{\mathcal{P}}'^*_{\hat{\mathbf{n}}'}\rangle - \langle \hat{\mathcal{P}}'_{\hat{\mathbf{n}}}\rangle \langle \hat{\mathcal{P}}'^*_{\hat{\mathbf{n}}'}\rangle \approx (F'^{-1}) F (F'^{-1}) \neq F^{-1}
\end{equation}
This implies that $\hat{\mathcal{P}}'_{\hat{\mathbf{n}}}$ constitutes a biased estimator, as we have discarded the information encoded in the smallest eigenvalues of the \ac{FIM} through truncation.
Nevertheless, this operation significantly reduces the estimation error for parameter $\mathcal{P}_{\hat{\mathbf{n}}}$.
We can use the covariance matrix of the clean map $\hat{\mathcal{P}}_{\hat{\mathbf{n}}}$ to measure the estimation error.
Since it is inconvenient to compare covariance matrices, in practice, we adopt the per-pixel standard deviation as a more convenient metric for evaluating estimation error,
\begin{equation}
    \sigma_{\rm pix} = \sqrt{F^{-1}_{\hat{\mathbf{n}}\hat{\mathbf{n}}'}}
\end{equation}
for unregularized estimators, and
\begin{equation}
    \sigma'_{\rm pix}  = \sqrt{ [(F'^{-1}) F (F'^{-1})]_{\hat{\mathbf{n}}\hat{\mathbf{n}}'}}
\end{equation}
for regularized estimators.

\section{Data simulation}\label{sec:simulation}
In this section, we provide a comprehensive description of our data simulation process, focusing specifically on pixel-based decomposition.
\subsection{Data generation}
We initiate our simulation by generating Gaussian \ac{SGWB}, whose covariance matrix $S_{IJ}^{\rm in}|_t(f)$ is computed through the summation of individual pixel contributions over the complete sky map,
\begin{equation}
    S_{IJ}^{\rm in}|_t(f) = \frac{4\pi}{N_{\rm pix}}\sum_{\hat{\mathbf{n}}}\gamma_{IJ}|_t(f, \hat{\mathbf{n}})\bar{H}^{\rm in}(f)\mathcal{P}^{\rm in}_{\hat{\mathbf{n}}},
\end{equation}
where $N_{\rm pix}$ denotes the number of pixels, the number of pixels directly determines two key factors: (i) sky localization resolution and (ii) computational load, with finer pixelization yielding better angular resolution at the expense of greater processing requirements. However, due to detector noise limitations, excessively fine angular resolution would be dominated by noise contamination. 
The indices $I$ and $J$ label distinct channels from different detectors. For the $A$, $E$, and $T$ channels specifically, this can be expressed as:
\begin{equation}
    \left( \begin{array}{ccc}
        AA' & AE' &AT'  \\
        EA' & EE' &ET'  \\
        TA' & TE' &TT'
    \end{array}
    \right)
\end{equation}
where the $AA'$ denotes the $A$ channels from different detectors. 

Considering the sampling frequency limitations of LISA and Taiji, we impose a frequency cutoff at 0.1 Hz and select two distinct spectral models $\alpha = 2/3$ and $\alpha = 3$ for signal injection. The key distinction lies in the spectral dependence: for $\alpha = 2/3$, the \ac{SNR} is predominantly contributed by low-frequency components, whereas for $\alpha = 3$, high-frequency components dominate the \ac{SNR}. 

For the noise, we generate a three-dimensional noise vector $n^{\rm in} = (n_A, n_E, n_T)$ based on the \ac{PSD} of each detector. We then compute the correlations among its components to construct the corresponding $3 \times 3$ noise covariance matrix $N^{\rm in}$. 
Under the assumption that the noise in different detectors is uncorrelated, the expectation value of $N_{IJ}$ is zero.
Notably, due to the effects of \ac{TDI}, Fig.~\ref{fig:noise PSD} exhibits several points where the noise PSD undergoes violent oscillations, including points reaching zero. These numerical artifacts would contribute to round-off errors 
which may introduce significant errors to the dirty map $X_{\hat{\mathbf{n}}}$ and \ac{FIM} $F_{\hat{\mathbf{n}}\hat{\mathbf{n}}'}$. Consequently, we exclude these frequency points in practical analysis to ensure data processing accuracy.

\subsection{Segment selection}\label{sec:segment}
The distinct operational modes of different detector networks lead to varying observational coverage: TianQin-LISA and TianQin-Taiji provide 6 months (3+3) of concurrent data annually, while Taiji-LISA offers full-year coverage. However, considering the detector response variations and maintaining signal stationarity assumptions, we employ short-term Fourier transforms to segment the data into intervals where both the signal can be approximated as stationary and the detector response remains approximately constant. 
For LISA and Taiji in their heliocentric orbits (1-year period), the detector response can be treated as constant within each data segment. However, TianQin's geocentric orbit (3.64-day period) exhibits rapid response variations, excessively long segments violate the constant-response assumption, while overly short segments would disproportionately sacrifice low-frequency information. 

Following Eq.~(\ref{eq:X_IJ}-\ref{eq:Fisher}), we treat the pair $\gamma_{IJ}|_t(f,\hat{\mathbf{n}})\gamma^*_{IJ}|_t(f,\hat{\mathbf{n}})$ as an integrated unit and define its all sky all frequency response error as:
\begin{equation}
\begin{aligned}
  {\rm{Error}}(t) =& \frac{1}{N_{f}N_{\rm pix}}\sum_{f}\sum_{i=1}^{N_{\rm pix}}w_i(f,\hat{\mathbf{n}}) \\
  &\frac{\left[\gamma_i(t,f,\hat{\mathbf{n}})\gamma_i^*(t,f,\hat{\mathbf{n}})-\gamma_{i}(t_{\rm mid},f,\hat{\mathbf{n}})\gamma^*_i(t_{\rm mid},f,\hat{\mathbf{n}})\right]}{\gamma_i(t_{\rm mid},f,\hat{\mathbf{n}})\gamma^*_i(t_{\rm mid},f,\hat{\mathbf{n}})}
\end{aligned}
\end{equation}
where $\gamma(t,f,\hat{\mathbf{n}})$ and $\gamma(t_{\rm mid},f,\hat{\mathbf{n}})$ represent the geometric factors at an arbitrary time and at the segment midpoint, respectively.
This differs from the previously defined $\gamma_{IJ}|_t(f,\hat{\mathbf{n}})$: $\gamma_{IJ}|_t(f,\hat{\mathbf{n}})$ approximates the response over a time segment, whereas $\gamma(t,f,\hat{\mathbf{n}})$ represents the instantaneous response at a specific time.
$N_{f}$ and $N_{\rm pix}$ is the number of frequency points and pixels. $w(f,\hat{\mathbf{n}})$ is a weighting function designed to mitigate the influence of pixels with weak responses, thereby preventing overestimation of the overall response variation error
\begin{equation}
  w_i(f,\hat{\mathbf{n}}) = \frac{\gamma_i(t_{\rm mid},f,\hat{\mathbf{n}})\gamma^*_i(t_{\rm mid},f,\hat{\mathbf{n}})}{ \sum_{i=1}^{N_{\rm pix}} \gamma_i(t_{\rm mid},f,\hat{\mathbf{n}})\gamma^*_i(t_{\rm mid},f,\hat{\mathbf{n}})}.
\end{equation}

\begin{figure}
    \centering
    \includegraphics[width=0.5\textwidth]{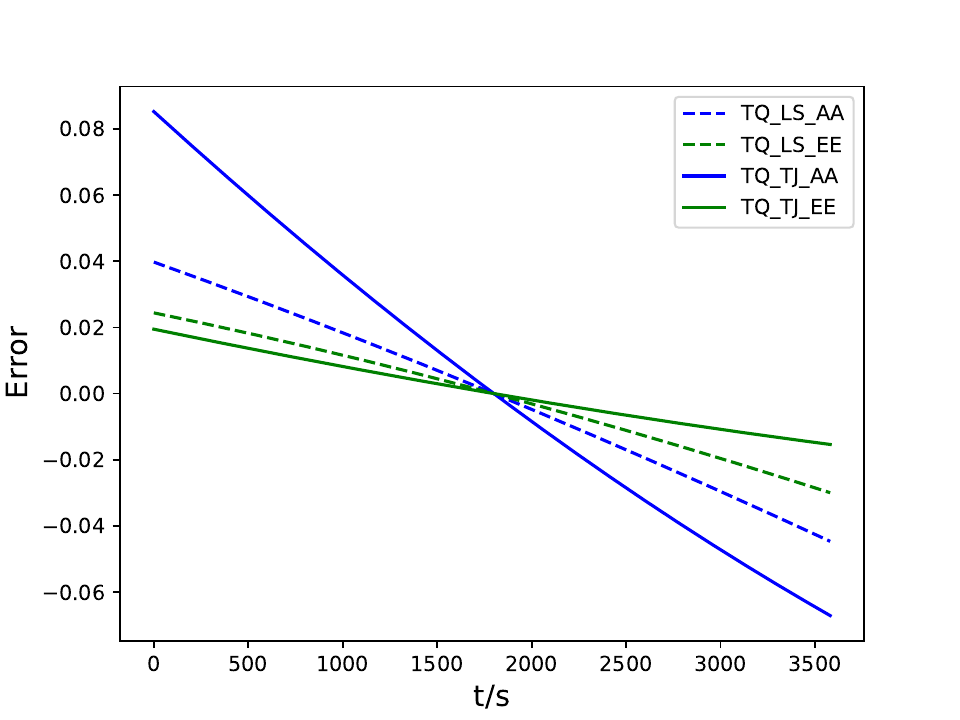}
    \caption{Relative error characteristics of the AA and EE correlation channels in the TQ+LS and TQ+TJ detector networks. The segment duration is set to 3600 seconds.}
    \label{fig:ORF_error}
\end{figure}

Figure~\ref{fig:ORF_error} presents the full-sky, all-frequency response relative errors for the TQ+LS and TQ+TJ configurations. 
In contrast, the TJ+LS network—comprising two heliocentric detectors—exhibits negligible response variations compared to the former pairs. 
We select one-hour intervals as our fundamental data segments, the minimum achievable frequency is $2.7 \times 10^{-4}$ Hz. 
Measurement results show the TQ+TJ AA channels attain peak errors of $15 \%$ at this segment. 
The assumption of constant response within these segments becomes invalid, while further reducing segment duration would significantly compromise low-frequency information retention. 
However, we note that the error within each segment evolves linearly over time. 
When comparing the response error at the segment midpoint versus the integrated segment error, the former remains significantly smaller than the absolute error across the full interval, which is smaller than $3\times10^{-3}$. 
For a one-hour data segment, approximating the geometric factor by its value at the segment midpoint introduces only negligible errors in our analysis. 
Accordingly, we adopt a one-hour duration as the standard segmentation length in our simulations.

\section{Results}\label{sec:result}
\subsection{Map recover}\label{sec:Map recover}
\begin{figure*}
    \centering
    \includegraphics[width=\textwidth]{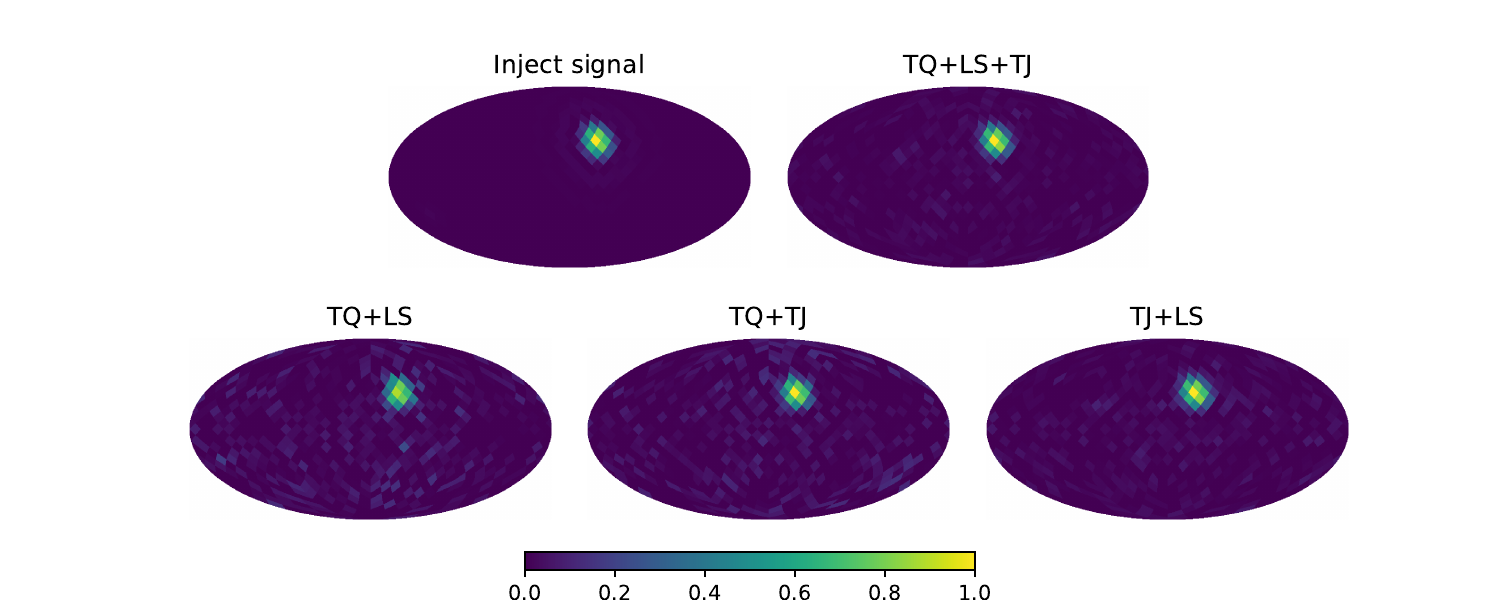}
    \caption{
    Normalized sky map recovery of gravitational-wave detector networks, assuming a power-law index of $\alpha = 3$.
    The top-left panel displays the injected \ac{SGWB} sky map in ecliptic coordinates, centered at $[\lambda, \beta] = [5\pi/3, \pi/6]$, which is expanded in spherical harmonics with coefficients retained up to $l \leq 16$.
    }
    \label{fig:alpha3 map}
\end{figure*}

\begin{figure*}
    \centering
    \includegraphics[width=\textwidth]{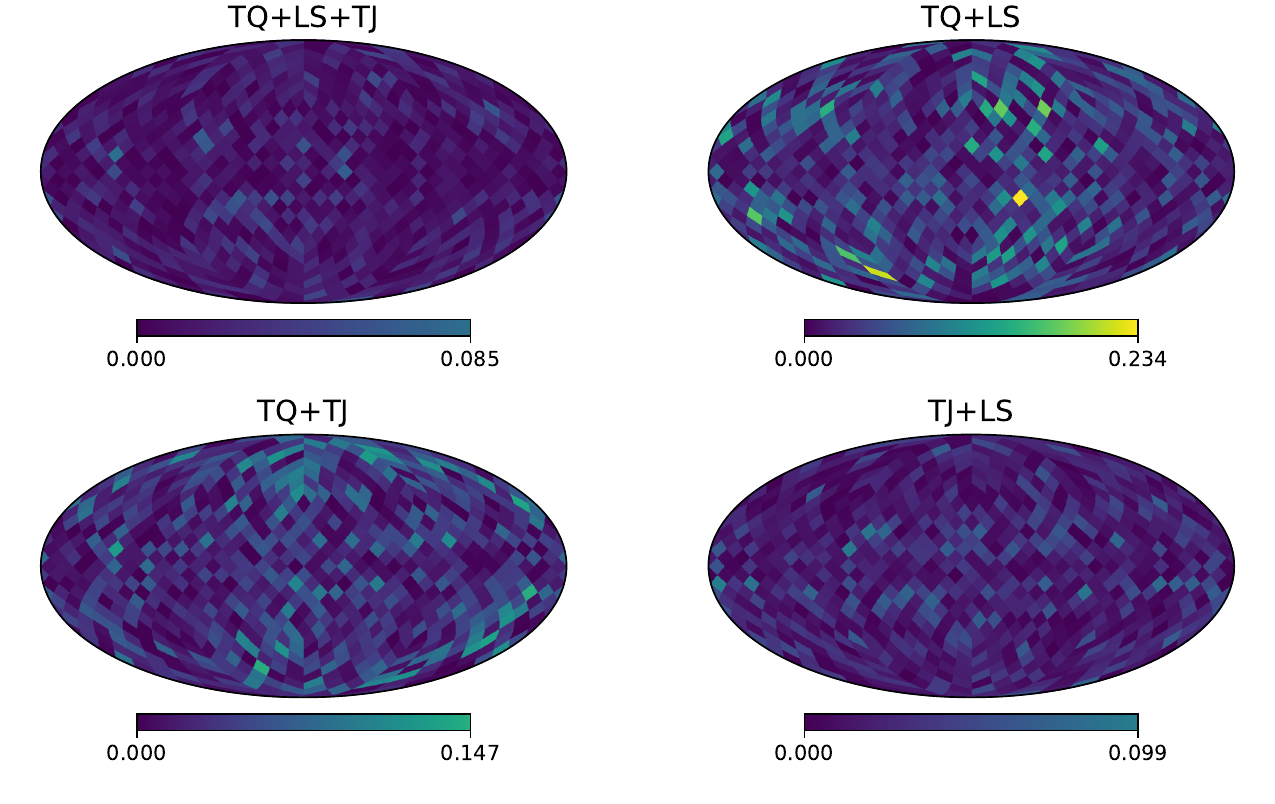}
    \caption{The residual maps between the injected sky map and the sky reconstructions obtained from different detector networks in Fig.~\ref{fig:alpha3 map}}
    \label{fig:map_residual}
\end{figure*}

\begin{figure}
    \centering
    \includegraphics[width=0.5\textwidth]{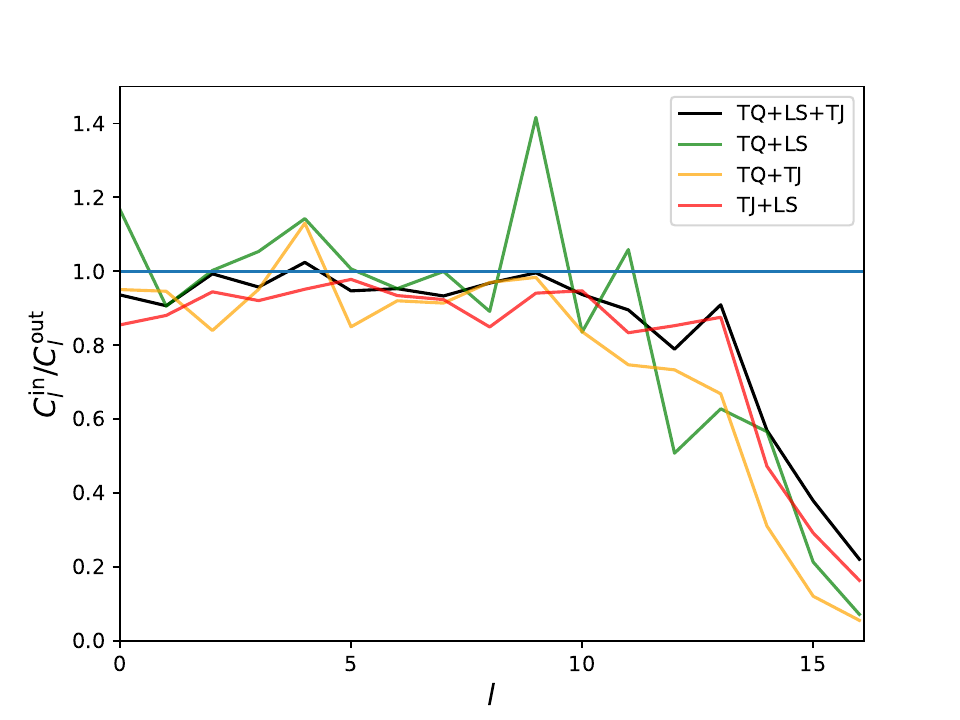}
    \caption{The angular power spectra $C_{l}$ ratio for sky map with $\alpha = 3$.  Here, the $C_l^{\rm in}$ denotes the angular power spectrum of the injected point source, as shown by `Inject signal' in Fig.~\ref{fig:alpha3 map}, while $C_l^{\rm out}$ represents the angular power spectra of the reconstructed sky maps from various detector combinations. A ratio of $C_l^{\rm in}/C_{l}^{\rm out}$	approaching unity indicates a more accurate reconstruction of the sky map.}
    \label{fig:cl alpha3}
\end{figure}
The process now shifts to map-making. We use the Gaussian, unpolarized point-source data simulated in the previous section to perform sky map reconstruction. All reconstruction results are derived from the A/E/T channels, with two spectral indices considered: $\alpha = 2/3$ and $\alpha = 3$.

Fig.~\ref{fig:alpha3 map} shows the sky map recovery performance of gravitational-wave detector networks, assuming a power-law index of $\alpha = 3$. The color scale has been normalized such that the brightest point in the injected sky map has a value of 1. 
The top-left panel displays the injected \ac{SGWB} sky map in ecliptic coordinates, centered at $[\lambda, \beta] = [5\pi/3, \pi/6]$, which is expanded in spherical harmonics with coefficients retained up to $l \leq 16$.

The network configurations work one year to achieve SNRs of: TianQin-LISA (TQ+LS): 25; TianQin-Taiji (TQ+TJ): 53; Taiji-LISA (TJ+LS): 44, and the total \ac{SNR} for the TQ + LS + TJ network is 73. We inject signals with the same amplitude into different detectors, with a reference frequency of $f_{\rm r} = 1 \mathrm{mHz}$ and $\Omega_{\rm r} = 6\times10^{-11}$. 

For a direct comparison, Fig.~\ref{fig:map_residual} presents the residual maps between the injected sky map and the sky reconstructions obtained from different detector networks in Fig.~\ref{fig:alpha3 map}
The combined TQ+LS+TJ detector network achieves optimal sky map recovery, whereas the TQ+LS configuration yields comparatively inferior results. 
This performance difference can be attributed to TianQin's unique sensitivity advantage in the 0.1–1 Hz frequency band. 
However, since both Taiji and LISA are subject to an upper frequency limit of 0.1 Hz due to sampling constraints, TianQin's high-frequency capability cannot be fully utilized in joint observations, thereby diminishing its relative advantage in such configurations.
Furthermore, Taiji's superior sensitivity also contributes to the enhanced performance observed in both the TQ+TJ and TJ+LS configurations.

A more familiar basis for describing directional information is the spherical harmonic multipole expansion. Fig.~\ref{fig:cl alpha3} uses Healpix to convert the injected and recovered sky maps from Fig.~\ref{fig:alpha3 map} into angular power spectra $C_{l}$ for direct comparison. Here, all the $C_l^{\rm in}$ denotes the angular power spectrum of the injected point source, as shown by `Inject signal' in Fig.~\ref{fig:alpha3 map}, while $C_l^{\rm out}$ represents the angular power spectra of the reconstructed sky maps from various detector combinations. A ratio of $C_l^{\rm in}/C_{l}^{\rm out}$	
approaching unity indicates a more accurate reconstruction of the sky map.
Since the injected point sources are expanded up to 
$l = 16$, we truncate the reconstruction at 
$l = 16$ accordingly. 
For the TQ+LS network, components with  $l <8 $ can be retained,  while the upper limit for TQ+TJ, TJ+LS, and TQ+LS+TJ network correspond to $l \sim 10-14$.

\subsection{Error analysis}
\begin{figure*}
    \centering
    \includegraphics[width=1\textwidth]{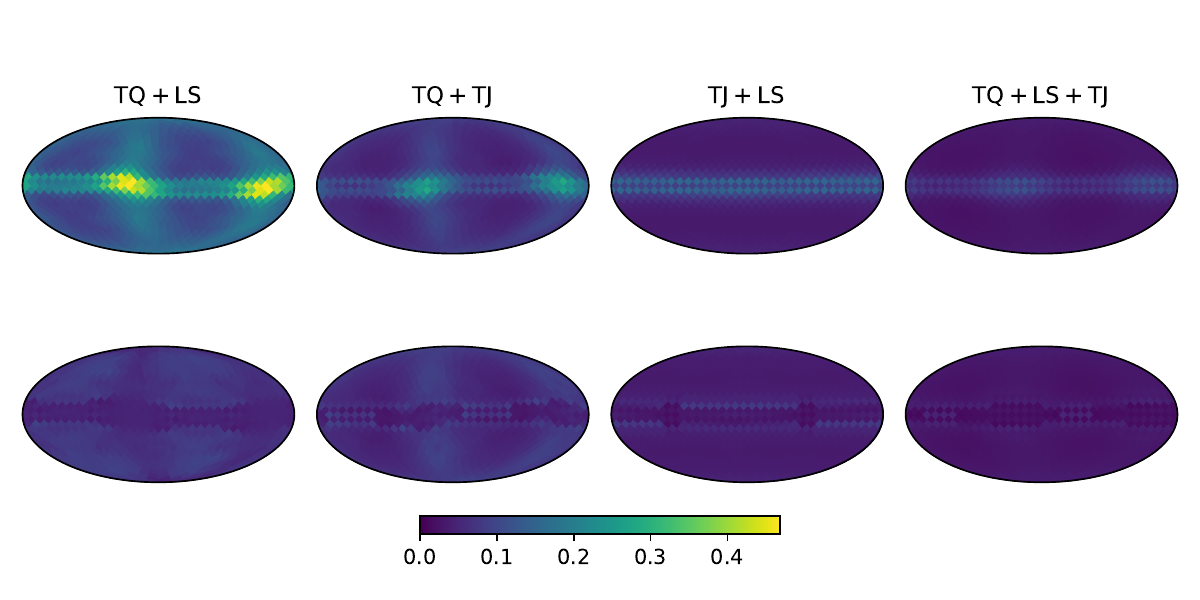}
    \caption{The per-pixel standard deviation maps for different detector networks. The top and bottom panels compare the results before and after SVD regularization, respectively. The regularization SVD reconstruction significantly reduces pixel-wise standard deviations across all networks, with the most pronounced improvements observed for TQ+TJ and TQ+LS configurations.}
    \label{fig:sigma_pix}
\end{figure*}
\begin{table}
    \centering
    \begin{tabular}{|c|c|c|}
    \hline
        Detector networks   & Before SVD & After SVD  \\
    \hline
       TQ+LS  & 0.468 &0.095 \\
       TQ+TJ  & 0.284 &0.095 \\
       TJ+LS  & 0.155 &0.079 \\
       TQ+LS+TJ &0.118 &0.049 \\
    \hline
    \end{tabular}
    \caption{The variation in maximum standard deviation values before and after SVD decomposition.}
    \label{tab:standard deviation}
\end{table}
The residuals in Fig.~\ref{fig:alpha3 map}-Fig.~\ref{fig:map_residual} are influenced by the deconvolution procedure employed during the map-making process. \lzy{I add the Fig.~\ref{fig:alpha3 map}}
The ill-conditioned nature of the \ac{FIM} leads to significant error amplification during its inversion, consequently increasing the residuals in the recovered sky map. By simulating 10 independent random realiszations with the same injection, we can obtain the standard deviation of recovered values on each pixel, and furthermore using this standard deviations to evaluate estimation error. 
Fig.~\ref{fig:sigma_pix} displays the per-pixel standard deviation maps for different detector networks. All results have been normalized. The top and bottom panels compare the results before and after SVD regularization, respectively. 
As clearly shown, the maximum errors for Taiji and LISA are concentrated near the ecliptic plane.
While for TianQin they are distributed along its orbital plane. This spatial distribution directly reflects the respective antenna response patterns of each detector~\cite{Li:2024lvt,Contaldi:2020rht}.
The regularization SVD reconstruction significantly reduces pixel-wise standard deviations across all networks. 
Table~\ref{tab:standard deviation} presents the variation in maximum standard deviation values across different detector combinations before and after SVD decomposition.

It is worth noting that the standard deviation of pixel estimates depends on the number of singular values retained during SVD. Retaining more singular values allows us to capture more high-order multipole information, but also leads to larger standard deviations. A balance between the two must be maintained.
In Table~\ref{tab:standard deviation}, the identical maximum standard deviation values for TQ+LS and TQ+TJ after SVD decomposition result from the more aggressive eigenvalue truncation strategy applied to TQ+LS network.
During implementation, TQ+LS retained only the top $58\%$ of eigenvalues, while TQ+TJ , TJ+LS, and TQ+TJ+LS retained the top $91\%$ of eigenvalues.
Had TQ+LS retained a comparable proportion of eigenvalues, the directional information of the signal would have been entirely obscured by deconvolution errors, while also introducing severe errors.

When working in the spherical harmonic basis, the transformation from pixel space introduces error propagation, making it difficult to directly derive the standard deviation of the angular power spectrum from the pixel basis. 
Furthermore, due to inherent numerical errors in coordinate transformations, a seemingly small residual in one basis (e.g., pixel space) may be significantly amplified when transformed into another basis (e.g., spherical harmonic space), leading to deviations substantially larger than initially anticipated~\cite{Contaldi:2020rht}.

\begin{figure}
    \centering
    \includegraphics[width=0.5\textwidth]{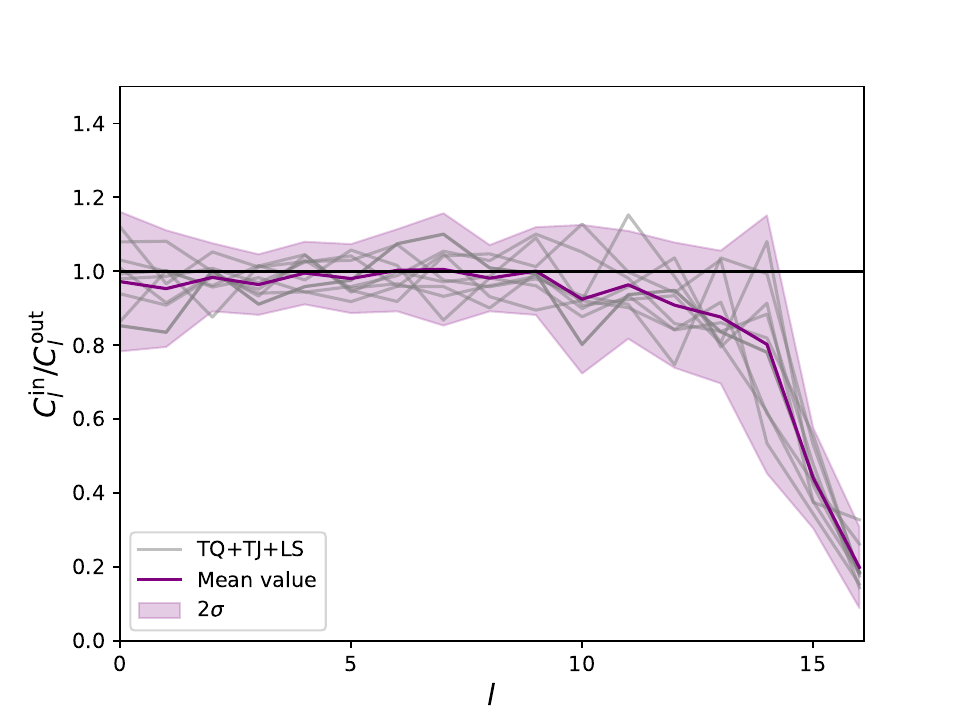}
    \caption{The $2 \sigma$ confidence interval of TQ+LS+TJ angular power spectra. The gray curves represent the angular power spectra reconstructed from ten individual realizations. The solid purple line denotes their mean value, and the shaded purple region indicates the $2 \sigma$ confidence interval derived from their sample variance.   }
    \label{fig:TQ_Tj_LS_errorbar}
\end{figure}

To address this, we performed multiple independent realizations and used the sample variance to represent the uncertainty in the angular power spectrum. 
Fig.~\ref{fig:TQ_Tj_LS_errorbar} presents the results of 10 independent realizations for the TQ+LS+TJ network. The gray curves represent the angular power spectra reconstructed from individual realizations, the solid purple line denotes their mean, and the shaded purple region indicates the $2 \sigma$ confidence interval derived from their sample variance. 
For multipole moments with $l \leq 10$, the angular power spectrum is reconstructed with high fidelity. 
Beyond $l > 10$, the reconstruction errors gradually increase, and for $l > 14$, the results deviate completely from the true values. 
This result is substantially better than those obtained with individual detectors such as LISA~\cite{Contaldi:2020rht} or TianQin~\cite{Li:2024lvt}. The multi-detector network, with its diverse spatial locations, provides observational data from multiple viewing angles, enabling significantly improved estimation of the stochastic gravitational-wave background's angular distribution.
\subsection{Other situation}
Beyond the case presented in Sec.~\ref{sec:Map recover}, we also systematically varied both the spectral shape and sky location of the injected signals to validate the robustness and reliability of our results.

\begin{figure}
    \centering
    \includegraphics[width=0.5\textwidth]{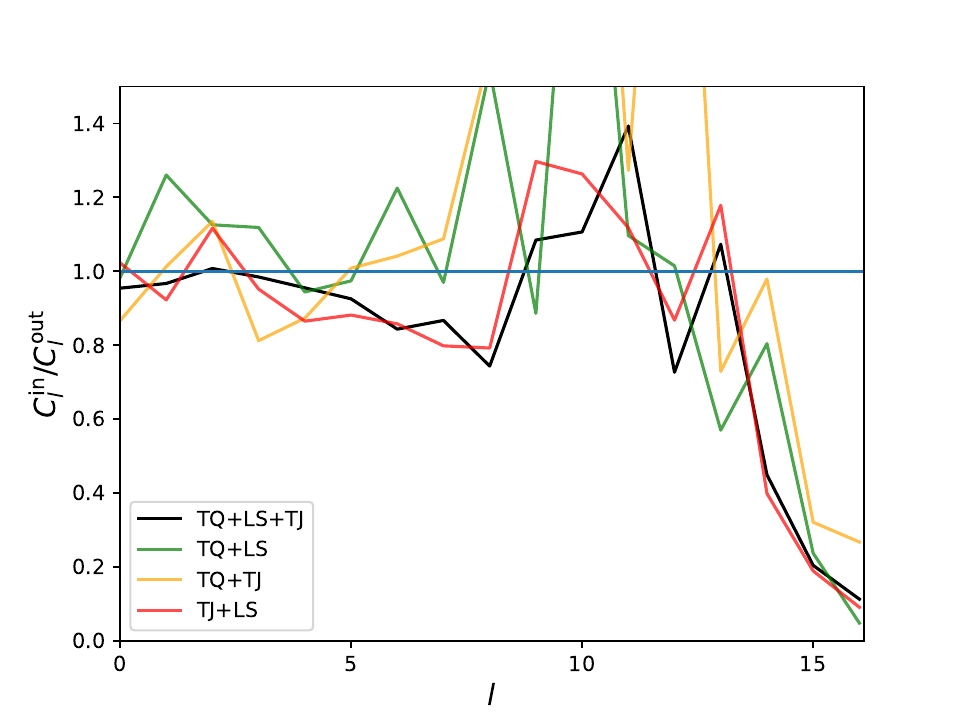}
    \caption{The angular power spectra $C_{l}$ ratio for sky map with $\alpha = 2/3$. Here, the $C_l^{\rm in}$ denotes the angular power spectrum of the injected point source, as shown by `Inject signal' in Fig.~\ref{fig:alpha3 map}, while $C_l^{\rm out}$ represents the angular power spectra of the reconstructed sky maps from various detector combinations.}
    \label{fig:Cl_compare}
\end{figure}

We began by changing the spectral index from $\alpha = 3$ to $\alpha = 2/3$, which shifts the primary frequency band contributing to the \ac{SNR} toward lower frequencies. 
Fig.~\ref{fig:Cl_compare} is the angular power spectra $C_{l}$ ratio for the sky map with $\alpha = 2/3$. We remain the injected point source parameters, as shown by `Inject signal' in Fig.~\ref{fig:alpha3 map}.
To ensure a consistent baseline for comparison, we adjust the individual detector sensitivities such that the lowest SNR among the network combinations remains comparable. Specifically, the network SNRs are set to 25 for TQ+LS, 36 for TQ+TJ, and 114 for TJ+LS, resulting in a total network SNR of 122.
We observe that the reconstruction results of all detector networks exhibit significant deviations at low multipole moments (low-$l$ values), with the TQ+LS and TQ+TJ configurations showing the most pronounced discrepancies.
This finding aligns with established research indicating that detector sensitivity to higher multipole moments decreases at lower frequencies~\cite{Taruya:2005yf,Taruya:2006kqa,Contaldi:2020rht,Li:2024lvt}.
Consequently, although the LS+TJ and TQ+LS+TJ networks achieve higher SNRs under the $\alpha = 2/3$ scenario compared to the $\alpha=3$ case, their reconstruction performance is ultimately degraded. 
Moreover, the reconstruction results of TQ+LS and TQ+TJ are even more degraded. In addition to the reasons mentioned above, this is further attributed to the relatively weaker sensitivity of TianQin in the low-frequency band.

\begin{figure*}
    \centering
    \includegraphics[width=\textwidth]{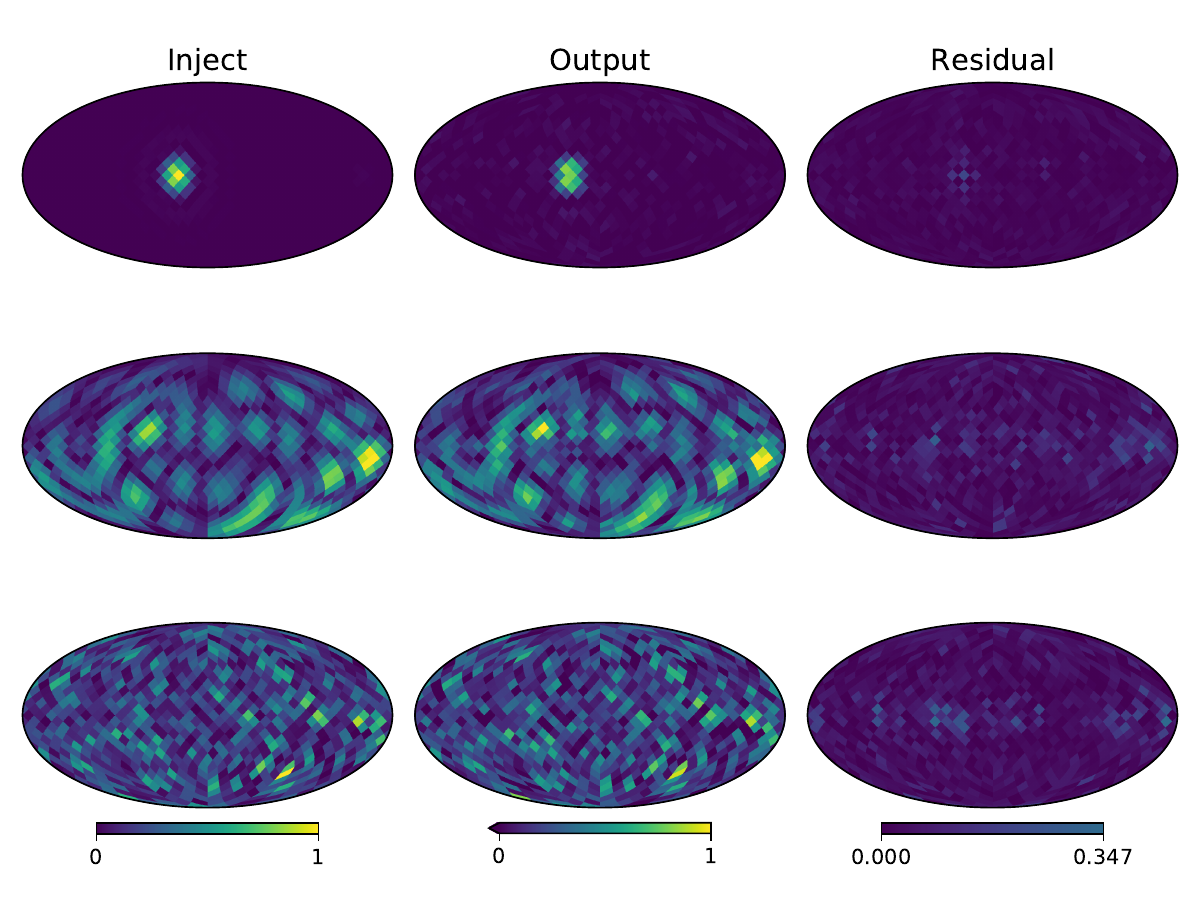}
    \caption{The point and random sky map recovery performance of TQ+LS+TJ networks. Assuming a power-law index of $\alpha = 3$. 
    The left column displays the injected sky maps: the top panel shows a point source, while the middle and bottom panels present random sky maps truncated at $l \leq 8$ and $l \leq 16$, respectively.
    The middle column displays the corresponding maps recovered using the TQ+LS+TJ network, while the right column presents the residual maps between the injected and output sky maps. Both the injected and output sky maps have been normalized, the two injected sky maps have comparable \acp{SNR} with the Fig.~\ref{fig:alpha3 map}. The maximum residual from top to bottom is 0.214, 0.312, and 0.347, respectively.}
    \label{fig:recover_Gau}
\end{figure*}

To further validate the robustness of our approach, we also modified the sky location of the injected maps. 
Firstly, we relocated the point source to the intersection of the ecliptic plane and TianQin's orbital plane, a region where the TQ+LS+TJ network exhibits minimal sensitivity.
Then, we generated two random sky maps truncated at $l \leq 8$ and $l \leq 16$, respectively.
Fig.~\ref{fig:recover_Gau} shows the point sources and random sky map recovery performance of TQ+LS+TJ networks, assuming a power-law index of $\alpha = 3$. 
The left column displays the injected sky maps: the top panel shows a point source, while the middle and bottom panels present random sky maps truncated at $l \leq 8$ and $l \leq 16$ respectively.
The middle column displays the corresponding maps recovered using the TQ+LS+TJ network, while the right column presents the residual maps between the injected and output sky maps. 
Both the injected and output sky maps have been normalized.
Compared with the scenario in Fig.~\ref{fig:alpha3 map}, here the injected signal is only differs in injection direction, with \acp{SNR} being comparable.
The maximum residual from top to bottom is 0.214, 0.312, and 0.347, respectively.

Compared to the residual map for the point source case in the top-left panel of Fig.~\ref{fig:map_residual}, the residuals in Fig.\ref{fig:recover_Gau} for the injected sky maps are substantially larger, with the highest deviations concentrated near the ecliptic plane. 
This outcome actually stems from two contributing factors. First, the concentration of larger residuals near the ecliptic plane is primarily due to the relatively lower response sensitivity of LISA and Taiji detectors in that sky area.
As shown in Fig.~\ref{fig:sigma_pix}, the pixels with the largest standard deviations in the TQ+LS+TJ configuration are concentrated along the ecliptic plane. When a point source is injected only in the upper-right region, the signal primarily lies in directions where the network has higher sensitivity, resulting in smaller overall residuals. However, when the sky map distribution is changed, especially when signals appear in low-sensitivity areas such as the ecliptic plane, the reconstruction quality will be significantly degraded in these regions.

On the other hand, the residuals from the random sky maps in Fig.\ref{fig:recover_Gau} (middle and bottom) are even larger than those generated by a point source located in the least sensitive region of the network (top).
This limitation originates from the inherent constraints of the pixel basis we selected for sky map reconstruction.
In the point sources case, only a small subset of pixels are activated (non-zero values), which significantly reduces the dimensionality of the parameter space in the deconvolution process and thereby greatly enhances parameter estimation accuracy. However, for randomly distributed sky maps, estimates are required for all pixels across the entire sky. This substantial expansion of the parameter space notably increases both the computational complexity and the uncertainty of the deconvolution procedure.

\section{Summary}\label{sec:summary}
In this paper, we have explored the map-making for the anisotropic SGWB with space-borne detector networks. We reconstruct the sky map using the maximum likelihood method based on simulated data with noise in pixel base. 
First, we consider a Gaussian, unpolarized point source, expand it in the spherical harmonic basis, and truncate the expansion at $l = 16$ to serve as the injected sky map. 
The results show that for a spectral index of $\alpha = 3$, considering the noise, the TQ+LS+TJ detector network can successfully reconstruct the angular power spectrum up to a multipole moment of $l \leq 14$ under ideal conditions.

To ensure the accuracy of our results, we subsequently computed the standard deviation of the estimates in both pixel space and spherical harmonic space. 
Analysis in pixel space reveals that the TQ+LS+TJ network exhibits significantly larger estimation errors near the ecliptic plane compared to other parts of the sky. This pattern originates from the detector configuration itself: LISA and Taiji, which contribute most to the network's signal-to-noise ratio, simultaneously display relatively weaker sensitivity around the ecliptic plane, directly leading to the increased estimation errors observed in this region for the combined network.
In spherical harmonic space, however, error propagation prevents direct derivation of the standard deviation of harmonic coefficients from pixel-space statistics. To address this, we performed 10 independent simulation realizations and used their sample variance as a robust estimate of the uncertainty in the harmonic domain.
The results indicate that the angular power spectrum is accurately reconstructed for multipole moments up to $l \leq 10$. Beyond this range, the reconstruction errors grow progressively with increasing $l$, and for $l > 14$, the results deviate completely from the true values.
Previous studies have shown that, when accounting for realistic noise, individual detectors can reconstruct the angular power spectrum up to a maximum multipole moment of $l<7$ for LISA~\cite{Contaldi:2020rht} and $l<4$ for TianQin~\cite{Li:2024lvt}.
Compared to a single detector, the detector network significantly improves the sky map reconstruction capability~\cite{Zhao:2024yau,Mu:2025gtg}.

An additional interesting question arises: if only two of the three detectors were launched and operated simultaneously, how substantial would the performance gap be compared to the full three-detector network?
Indeed, the advantages of a multi-detector network are manifested in three key aspects. First, the increased observational data enhances the \ac{SNR}. Furthermore, the diverse spatial locations of different detectors provide complementary viewing angles, reducing blind spots in sky map and significantly diminishing the ill-conditioning of the Fisher information matrix, thereby improving sky map reconstruction. 
Most importantly, such a network configuration reduces the reliance on precise knowledge of individual detector noise parameters, which remains an open challenge, despite recent efforts that have begun to address this issue~\cite{Baghi:2023qnq,Muratore:2023gxh,Santini:2025iuj}.
For detector networks, they do not require prior knowledge of the noise properties, which greatly enhances the robustness and reliability of the results. 
Therefore, the absence of any future detector would reduce the \ac{SNR} of our data and expand blind regions in the sky map, leading to some reduction in parameter estimation accuracy. Nevertheless, the multi-detector network would still yield significantly more reliable results than those achievable with a single detector.


Finally, in this work, some idealizations still have been retained, 
we assume the the signal's amplitude and direction are separable, which may not hold for certain anisotropic \ac{SGWB}~\cite{Cusin:2017fwz,ValbusaDallArmi:2022htu,Chung:2022xhv,LISACosmologyWorkingGroup:2022kbp,Chowdhury:2022pnv,Cusin:2022cbb,Cruz:2024svc,Heisenberg:2024var}. Besides,
for simplicity, we model the noise as Gaussian and identical across all laser links, whereas in reality, it is significantly more complex and heterogeneous~\cite{Hartwig:2023pft,Wang:2024ovi,Alvey:2024uoc,Kume:2024xvh,Du:2025xdq}. 
Whether sky map reconstruction would be significantly affected under more realistic noise conditions remains an open question and warrants further investigation.
\begin{acknowledgments}
This work has been supported by the National Key Research and Development Program of China (No. 2023YFC2206704, No. 2020YFC2201400), the Natural Science Foundation of China (Grants No. 12173104), the Natural Science Foundation of Guangdong Province of China (Grant No. 2022A1515011862), and the Guangdong Basic and Applied Basic Research Foundation(Grant No. 2023A1515030116). 
Z.C.L. is supported by the Guangdong Basic and Applied Basic Research Foundation (Grant No. 2023A1515111184). 
\end{acknowledgments}

\normalem
\bibliographystyle{apsrev4-1}
\bibliography{ref}
\end{document}